\documentclass{aa}  
\usepackage{graphicx}
\usepackage{color}

\begin{document}
   \title{Magnetic field structure in single late-type giants: The effectively single giant V390 Aur\thanks{Based on data obtained using the T\'elescope Bernard Lyot at Observatoire du Pic du Midi, CNRS and Universit\'e Paul Sabatier, France.}}


   \author{ R. Konstantinova-Antova\inst{1,2}, M. Auri\`ere\inst{2}, P. Petit\inst{2}, C. Charbonnel\inst{3,2}, S. Tsvetkova \inst{1}, A. L\`ebre \inst{4}, and R. Bogdanovski \inst{1} }

   \offprints{R. Konstantinova-Antova}

   \institute{Institute of Astronomy and NAO, Bulgarian Academy of Sciences, 72 Tsarigradsko shose, 1784 Sofia,
             Bulgaria\\
              \email{renada@astro.bas.bg}
\and
IRAP; UMR 5277; CNRS and Universit\'e de Toulouse, 14, Avenue Edouard Belin, F-31400 Toulouse, France 
\and
Geneva Observatory, University of Geneva, 51, Chemin des Maillettes, 1290 Versoix, Switzerland
\and
LUPM - UMR5299; CNRS and Universit\'e de Montpellier 2, Place E. Bataillon, F-34095 Montpellier cedex 05, France\\}

   \date{Received ; }

\abstract 
{}
{We have studied the active giant V390 Aur using   
spectropolarimetry to obtain direct and simultaneous 
measurements of the magnetic field and the activity indicators in 
order to get a precise insight of its activity. }
{We used the spectropolarimeter NARVAL at the Bernard Lyot Telescope
(Observatoire du Pic du Midi, France) to obtain a series of Stokes I and 
Stokes V profiles. The Least Square deconvolution (LSD) technique was applied to detect the 
Zeeman signature of the magnetic field in each of our 13 observations 
and to measure its longitudinal component. We could also monitor the $CaII K\&H$ and 
IR triplet, as well as the H$_\alpha$ lines which are activity indicators. 
In order to reconstruct the magnetic field geometry of V390 Aur, we applied the Zeeman Doppler Imaging (ZDI) inversion method and present a map for the magnetic field. Based on the obtained spectra, we also refined the fundamental parameters of the star and the Li abundance.}
{The ZDI revealed a structure in the radial magnetic field  consisting of a polar magnetic spot of positive polarity and several negative spots at lower latitude. A high latitude belt is present on the azimuthal field map, indicative of a toroidal field close to the surface.  It was found that the photometric period cannot fit the behaviour of the activity indicators formed in the chromosphere. Their behaviour suggests slower rotation compared to the photosphere, but our dataset is too short to be able to estimate the exact periods for them.

All these results can be explained in terms of an $\alpha-\omega$ dynamo operation, taking into account the stellar structure and rotation properties of V390 Aur that we study using up to-date stellar models computed at solar metallicity. The calculated Rossby number also points to a very efficient dynamo.}
{}{}

   \authorrunning{Konstantinova-Antova et al.}
   \titlerunning{Magnetic field structure in V390 Aur}
   \maketitle 

\section{Introduction}

The G8 III star V390 Aur = HD33798 is already known for its magnetic activity. A modulation of the photometric light curve due to spots is observed by Spurr \& Hoff (\cite{sh}) and a period of 9.825 days that is considered the rotational period of the star is determined by Hooten \& Hall (\cite{hh}). It has been classified as a 
chromospherically active giant by Fekel \& Marshall (\cite{fm}) on the basis of its $CaII K$ emission and fast rotation ($vsini=29\mathrm{\,km.s^{-1}}$).
Later, enhanced X--ray emission (H\"unsch et al. \cite{hunsch}, Gondoin \cite{gondoin99}, Gondoin \cite{gondoin03}) and optical flares 
(Konstantinova-Antova et al. 2000, 2005) were detected for this star. Recently, Konstantinova-Antova et al. (2008) reported direct detection of magnetic field using the spectropolarimeter NARVAL at the 2m Telescope Bernard Lyot at Pic du Midi Observatory, France. 

According to the literature V390 Aur has a mass of about 1.8--2~$M_{\sun}$ and is evolving along the base of the red giant branch (RGB) (Gondoin \cite{gondoin03}, Konstantinova-Antova et al. 2009). It is the primary of a wide multiple physical system, ADS 3812, and it was shown that the giant   
could be considered effectively single with respect to its activity, i.e. synchronization plays no role in its fast rotation and activity (Konstantinova-Antova et al. 2008).

Here we present the first map of the surface magnetic field structure for this giant star applying the Zeeman Doppler Imaging (ZDI) technique (Donati\& Brown 1997, Donati et al. 2006a). Section 2 describes our observations and methods for data processing. Section 3 gives our results from the ZDI and the behaviour of the activity indicators. Based on up-to-date stellar evolution models we estimate in Section 4 the mass and evolutionary status of V390 Aur and determine its Rossby number as an indicator of the dynamo efficiency. Section 5 and 6 are for the discussion and conclusions, respectively.

\section{Observations and data processing}

The observations of V390 Aur were performed at the 2-m
Bernard Lyot Telescope (TBL) of Pic du 
Midi Observatory with the new generation spectropolarimeter NARVAL (Auri\`ere \cite{auriere03}) which is a copy of the instrument ESPaDOnS installed at the CFHT at the end of 2004 
(Donati et al. 2006b). NARVAL is a fiber--fed echelle spectrometer 
allowing the whole spectrum from 370 nm to 1000 nm to be recorded in each exposure, in 40 orders aligned in the CCD frame by 2 cross-disperser prisms.
We used it in polarimetric mode with a spectral resolution of 
about 65000. Stokes I (unpolarised) and Stokes V (circular polarization) 
parameters were obtained by four sub-exposures, where in between each of them the retarders and Fresnel rhombs were rotated in order to exchange the beams 
in the instrument and to reduce the spurious polarization signatures (Semel et al. 1993).

Thirteen observations were obtained in the period 14 -- 30 September, 2008 (Table\ref{table:1}).  
The extraction of the spectra was performed using Libre-ESpRIT 
(Donati et al. \cite{donati97}),
a fully automatic reduction package installed at TBL. For the Zeeman analysis, least-square deconvolution analysis 
(LSD, Donati et al. \cite{donati97}) was applied to all observations. We used a mask calculated for an effective 
temperature of $5000\,K$, $\log g =3.0$ and a microturbulence of $2.0\mathrm{\,km.s^{-1}}$, consistent with physical parameters 
given by Gondoin (\cite{gondoin03}). For the case of V390 Aur, the method enabled us to average about 12,300 lines and to get 
Stokes I and Stokes V profiles with a significantly improved signal--to--noise ratio (S/N). 
A significant Zeeman Stokes V signal was detected for each observation. The null spectrum given by the standard 
procedure (Donati et al. \cite{donati97}) was also examined, which showed no signal. We 
then computed the longitudinal magnetic field B$_{l}$ in G, using the 
first-order moment method (Donati et al. \cite{donati97}, Rees and Semel \cite{rees79}). To determine B$_{l}$ values, 
we integrated the LSD profiles between -17 and $+66\mathrm{\,km.s^{-1}}$, which reduced the significant blending by the profiles of the faint stellar companion of V390 Aur (which is a SB2 star, Konstantinova-Antova et al. 2008). These profiles are in some of our spectra.

\begin{figure}
   \centering
   \includegraphics[width=8cm, angle=0]{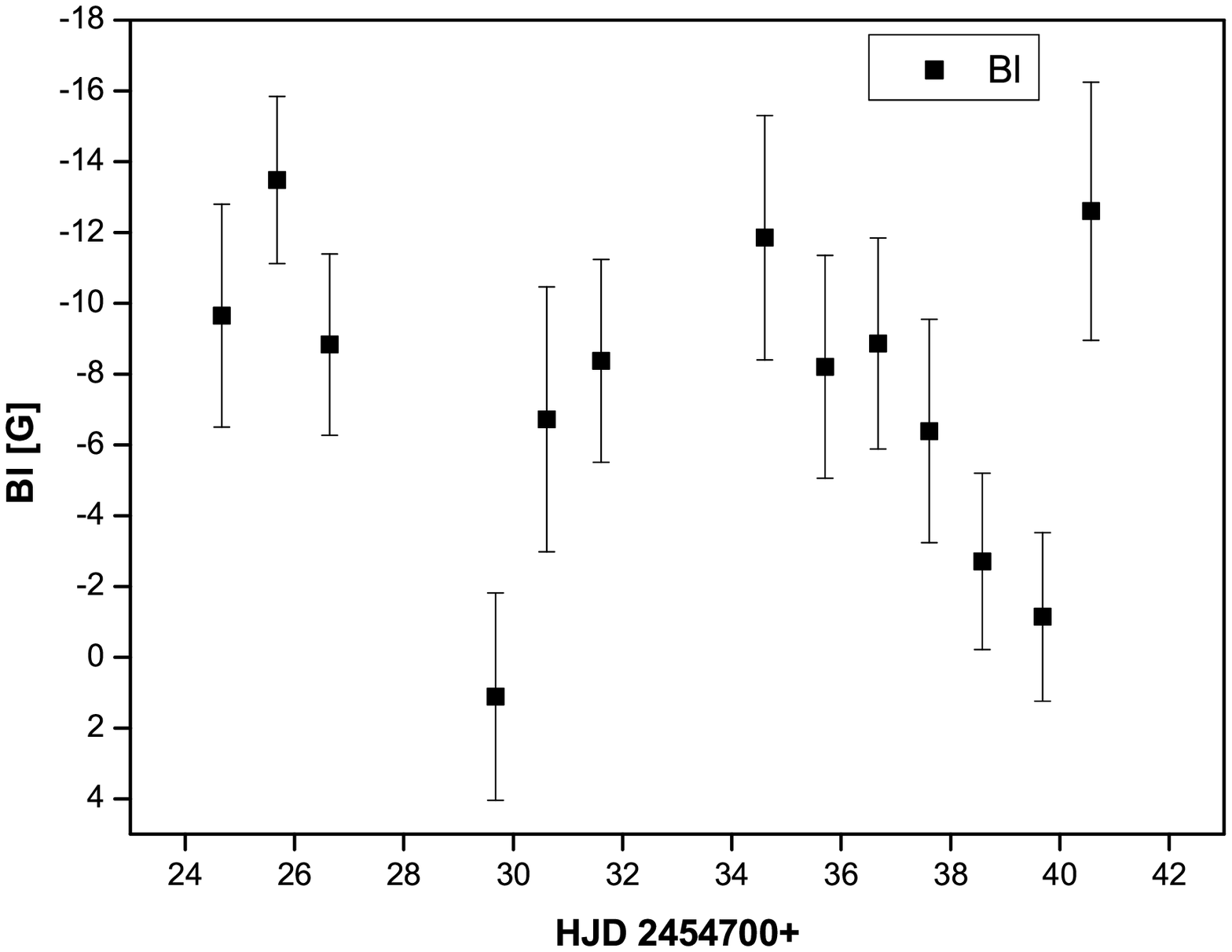}
   \includegraphics[width=8cm, angle=0]{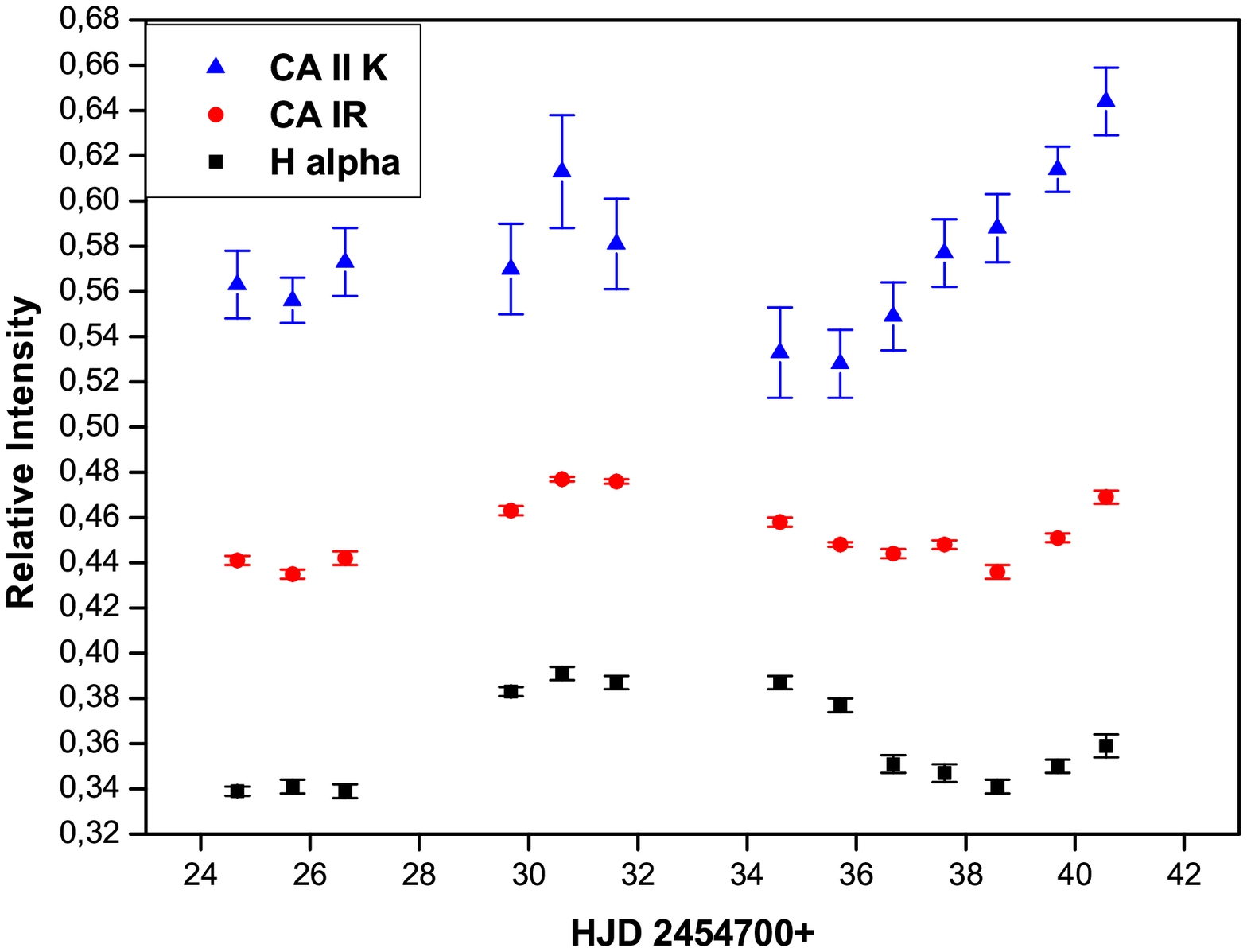}
      \caption{Variability of B$_{l}$ (top), $CaII~K$, $CaII~IR~854.2\, \mathrm{nm}$ and $H_\alpha$ (bottom) for V390 Aur.}
         \label{Figure1}    \end{figure}

The activity of the star for the same period has been monitored with 
measurements of the relative intensity regarding the continuum (R$_c$) for the line--activity indicators $CaII~IR~854.2\, \mathrm{nm}$ and  
$H_\alpha$.  For $CaII~K$ the relative intensity of the emission core I/I(395\,nm) is measured. The S/N is greater than 50 for the $CaII~K\&H$ emission cores and greater than 500 for the rest of the spectral lines mentioned above. The B$_{l}$ and activity indicators behaviour is presented in Table\ref{table:1} and in Fig.\ref{Figure1}.

\begin{table}
\caption{Data for activity indicators and B$_{l}$.}             
\label{table:1}      
\centering                          
\begin{tabular}{c c c c c c c}        
\hline\hline                 
Date&  Phase& CaII K & H$_\alpha$& CaII IR &B$_l$& $\sigma$ \\
     &     &        &           &         &Gauss& Gauss\\
\hline                        
14 Sep 08&0.00&0.56&0.339 & 0.441&$-$10&3 \\ 
15 Sep 08&0.10&0.55&0.341 & 0.435&$-$13&2 \\
16 Sep 08&0.20&0.57&0.339 & 0.442&$-$9 &3 \\
19 Sep 08&0.51&0.57&0.383 & 0.463&1  &3 \\
20 Sep 08&0.60&0.61&0.391 & 0.477&$-$7 &4 \\
21 Sep 08&0.705&0.58&0.387& 0.476&$-$8 &3 \\
24 Sep 08&0.01&0.53&0.387 & 0.458&$-$12&3 \\
25 Sep 08&0.12&0.52&0.377 & 0.448&$-$8 &3 \\
26 Sep 08&0.22&0.55&0.351 & 0.444&$-$9 &3 \\
27 Sep 08&0.32&0.58&0.347 & 0.448&$-$6 &3 \\
28 Sep 08&0.42&0.59&0.341 & 0.436&$-$3 &2 \\
29 Sep 08&0.53&0.61&0.350 & 0.451&$-$1 &2 \\
30 Sep 08&0.62&0.64&0.359 & 0.469&$-$13&4 \\
\hline                                   
\end{tabular}
Note: $R_c$ values are given for $H_\alpha$ and $CaII~854.2\,\mathrm{nm}$. For $CaII~K$ emission core I/I(395\,nm) is measured. B$_l$ values for the magnetic field and their accuracy are given in Gauss. Phase is for 9.825d photometric period, 
considered as the rotational period of V390 Aur.
\end{table}

In addition, we performed ZDI (Donati et al. 2006a) for V390 Aur, using all these observations. In spite of a large $vsini$ value that makes V390 Aur a possible target for classical Doppler Imaging, we do not present here a photospheric map of brightness inhomogeneities, because of the occasional Stokes I contamination by the double-lined stellar system associated with the giant. Without any accurate orbital parameters available for this system, no clean removal of its Stokes I signatures can be performed following the approach of Donati (\cite{donati99}). The primary is the only detected contributor to the Stokes V line profiles, so that other components of the system do not affect the magnetic map reconstruction.

\section{The magnetic field geometry of V390 Aur and the behaviour of the activity indicators}

\subsection{Zeeman Doppler Imaging}
In order to reconstruct the magnetic field geometry of V390 Aur based on modelling of the rotational modulation of the Stokes V profiles, we applied the ZDI inversion method (Donati \& Brown 1997). The inversion code we used is described in Donati et al. (2006b) and enables the split of the magnetic field in poloidal and toroidal components.

To calculate the rotation phases of our observations, we took as rotational period the photometric one of 9.825 days, $vsini=29\mathrm{\,km.s^{-1}}$, inclination angle of the axis of rotation $56^\circ$ (Konstantinova-Antova et al. 2008), and a limb darkening coefficient of 0.75. We also assumed a constant radial velocity of $23.4\mathrm{\,km.s^{-1}}$. Taking into account the complex structure of the Stokes V profiles, we extend the order of the spherical harmonics modes to $\ell \leq 25$, since no improvement in the fit to the data is achieved by increasing further the maximum allowed value for $\ell$. 

Prior to reconstructing the final magnetic map, we have estimated the surface differential rotation of V390 Aur, using the method proposed by Petit et al. (2002). We assume a simple latitude dependence of the rotational shear in the form $\Omega(l) = \Omega_{\rm eq} - \Delta \Omega \sin^2 l$, where $\Omega(l)$ is the rotation rate at latitude $l$, $\Omega_{\rm eq}$ the rotation rate of the equator and $\Delta \Omega$ the difference in rotation rate between the polar region and the equator. The best magnetic model is achieved for differenial rotation parameters equal to $\Omega_{eq} = 0.652\pm0.002\mathrm{\,rad.d^{-1}}$ and $\Delta \Omega = 0.048\pm0.007\mathrm{\,rad.d^{-1}}$, with a reduced $\chi^2$ equal to 1. The value of the photometric period lies between the equatorial and polar periods estimated here and corresponds to the rotation period obtained around a latitude of $30^\circ$, which is the latitude seen orthogonally by the observer, assuming an inclination angle of $56^\circ$.

\begin{figure}
   \centering
\includegraphics[width=6.3cm, angle=270]{v390aur_V_renada3_1.eps}
   \includegraphics[width=11cm, height=12cm, angle=0]{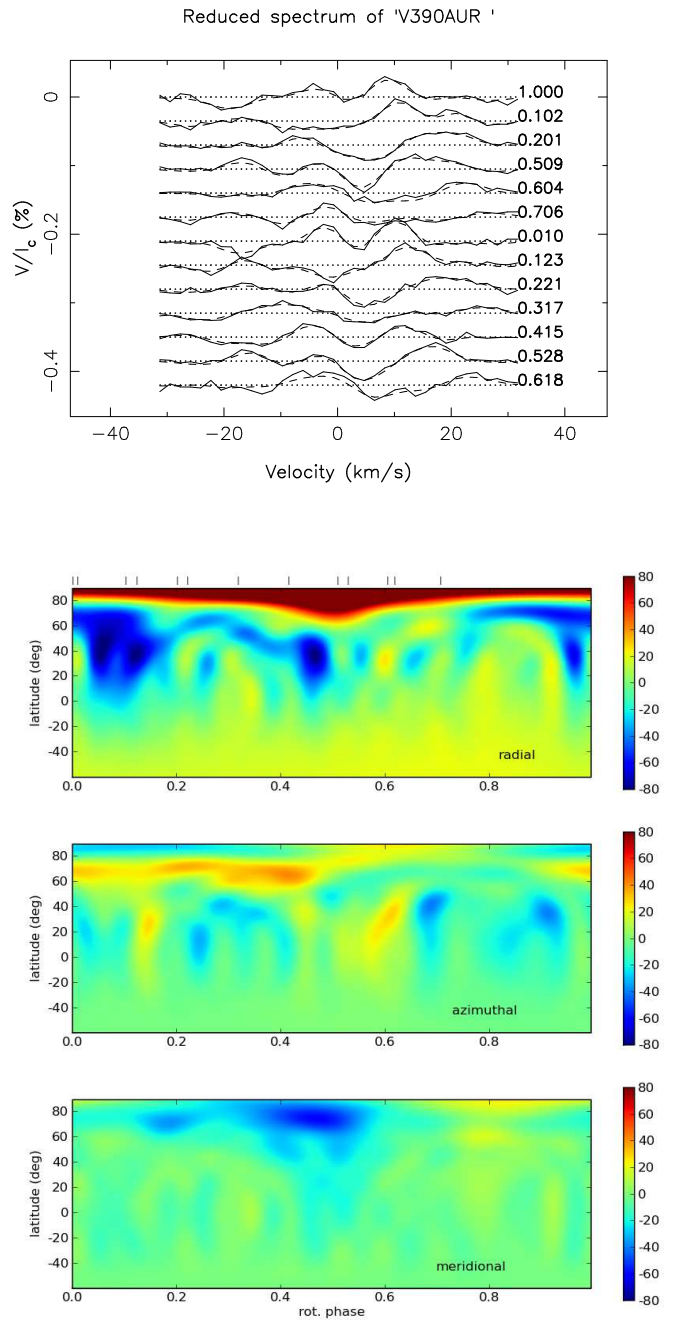}
      \caption{Upper pannel: LSD Stokes V profiles (solid lines) and their fit (dashed lines) after a correction of the mean radial velocity of the star. For display purposes the profiles are shifted vertically and the dotted horizontal lines indicate the zero level. Lower pannels: Radial, azimuthal and meridional maps for the reconstructed magnetic field for V390 Aur. The vertical ticks in the upper part of the radial map show the phases when there are observations.}
         \label{Figure2}
   \end{figure}

The result of our modelling is presented in Fig.\ref{Figure2}. In this figure, the upper pannel displays the Stokes V profiles fitted with the model, and the three other panels show the surface magnetic field distribution for V390 Aur as maps of the radial, azimuthal and meridional components of the magnetic field. 

Our map of the radial magnetic field reveals a strong polar spot of positive polarity and several groups of spots of negative polarity situated at lower latitudes. The azimuthal component of the field presents a middle to high latitude belt. We reconstruct about 80 percent of the surface magnetic energy in the poloidal magnetic component where the dipole structure dominates.

\subsection{B$_{l}$ and activity indicators}

\begin{figure}
   \centering
   \includegraphics[width=8cm, angle=0]{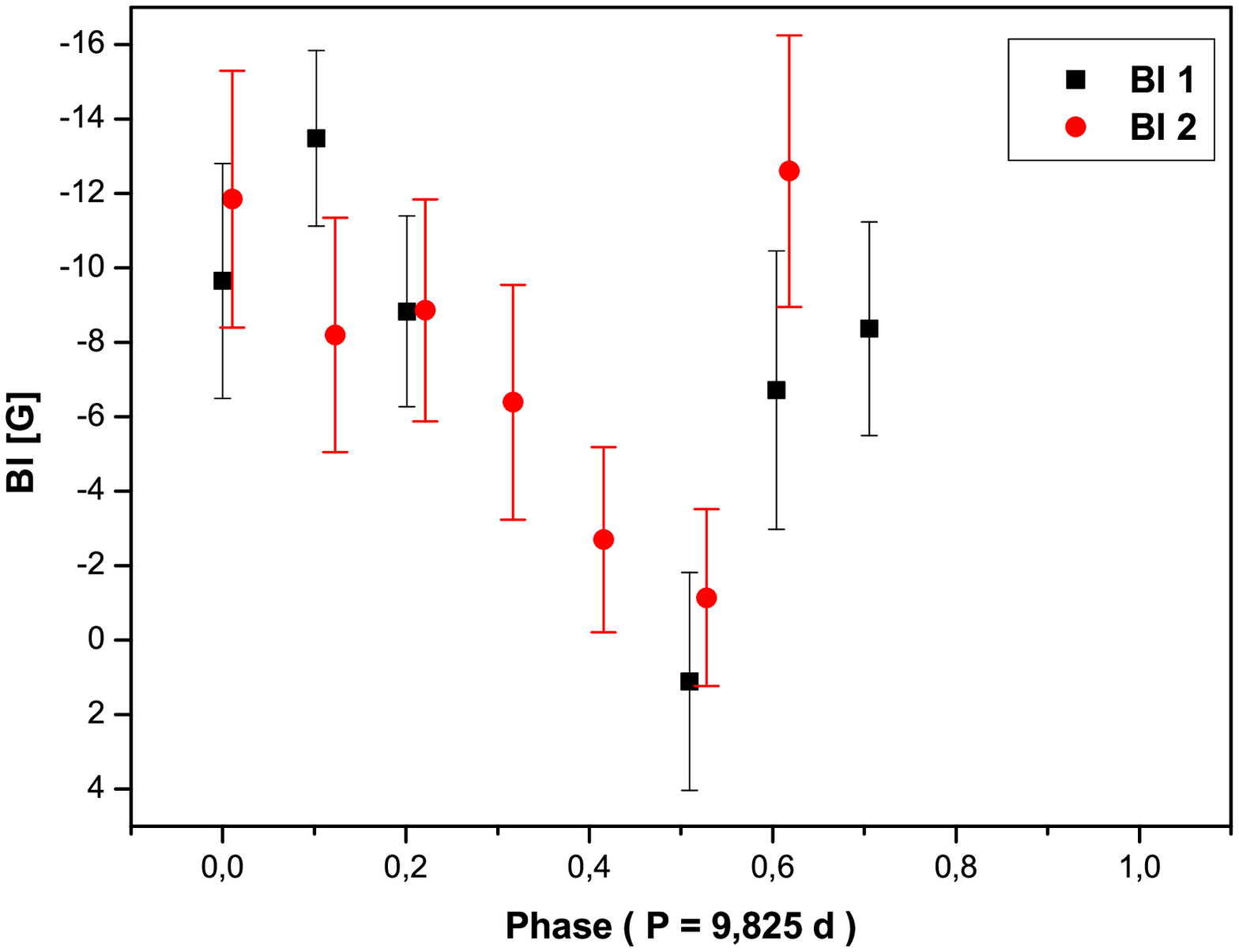}
   \includegraphics[width=8cm, angle=0]{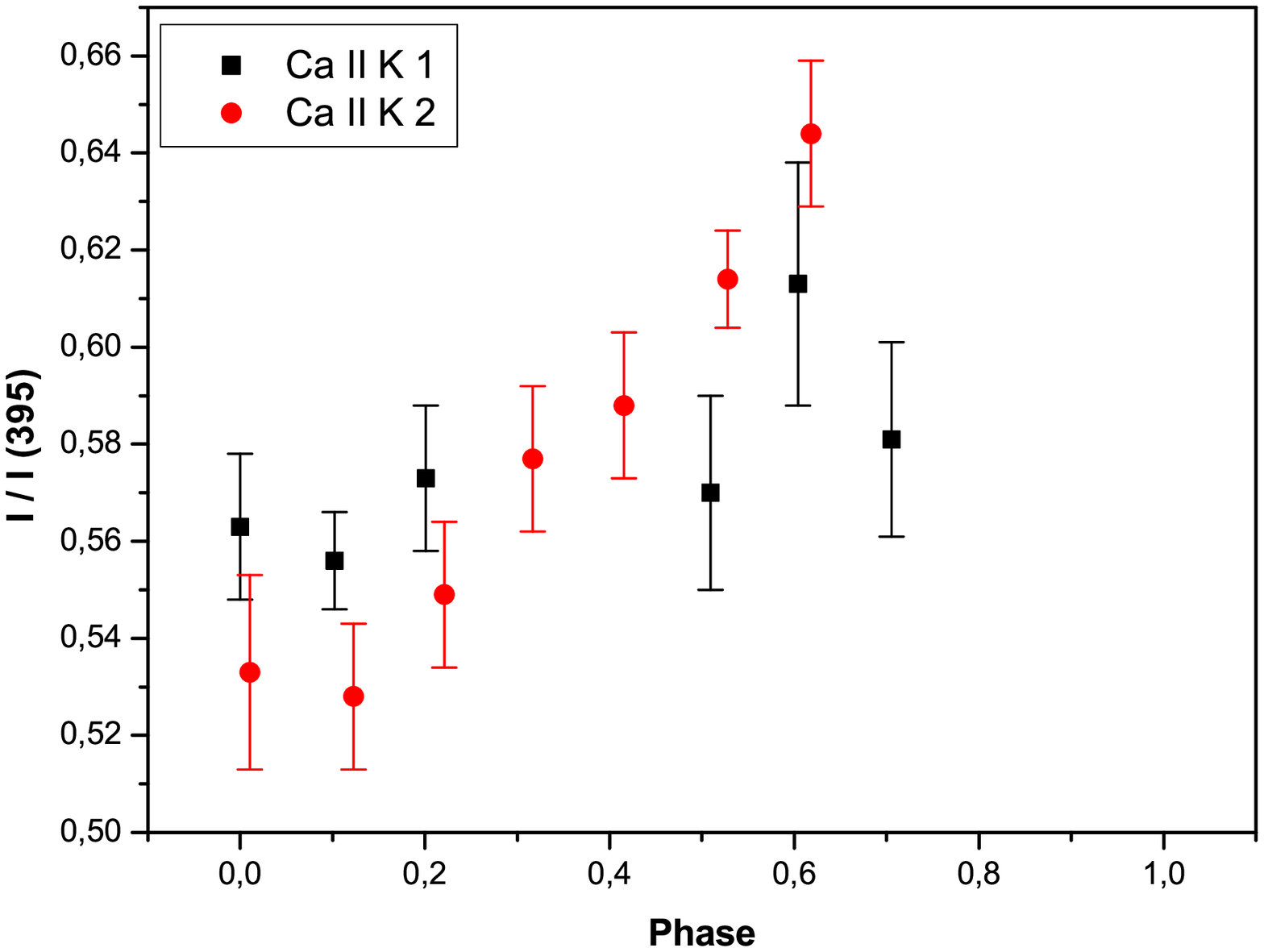}
   \includegraphics[width=8cm, angle=0]{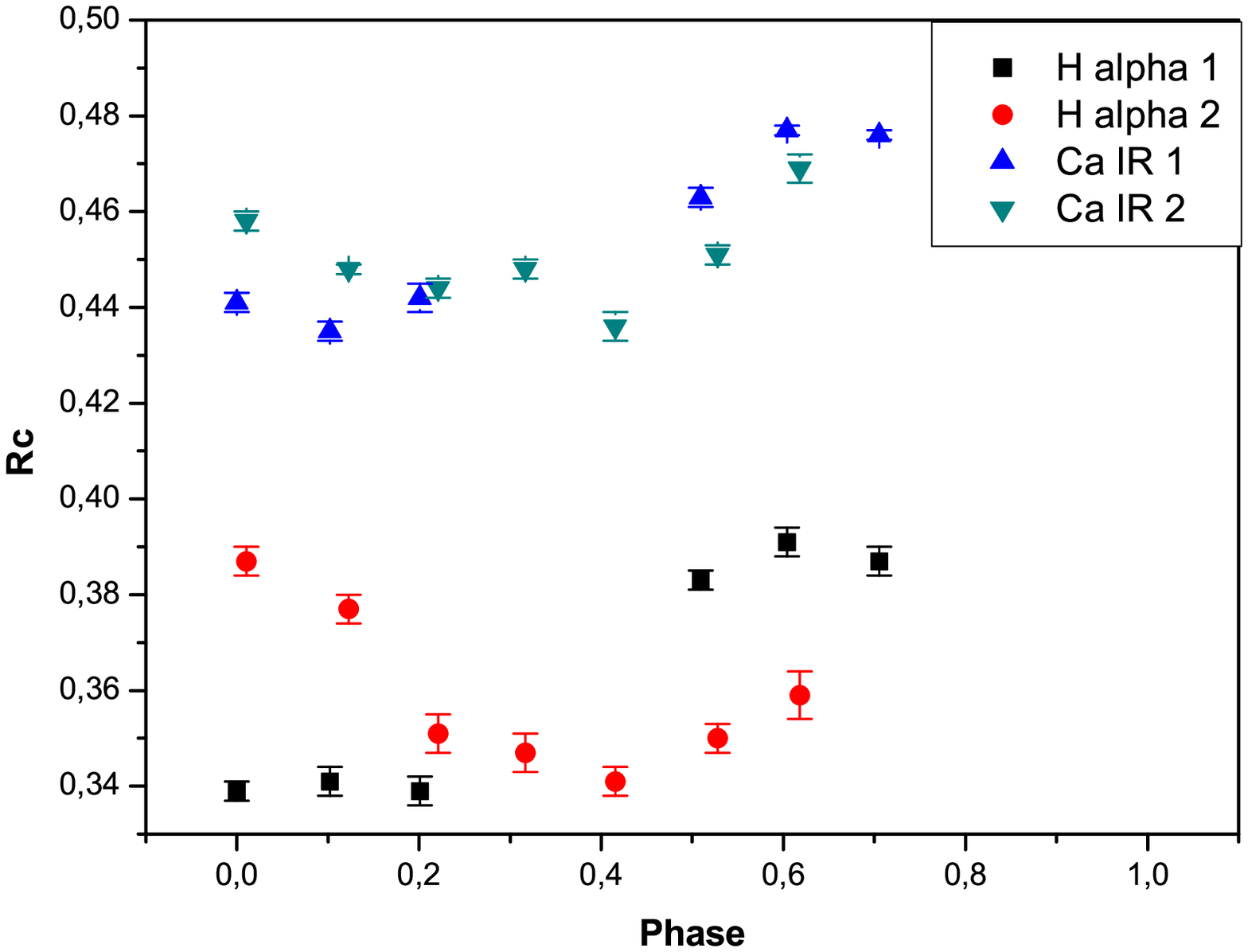}
      \caption{B$_{l}$, $CaII~K$, $H_\alpha$ and $CaII~IR~854.2\,\mathrm{nm}$ behaviour for V390 Aur. Phase is for the photometric period of 9.825\,d. The different colors in each of them correspond to a different rotation (see the legend on each pannel).}
         \label{Figure3}    
    \end{figure}

Figure \ref{Figure3}, upper panel, shows the variations of B$_{l}$  with respect to the photometric phase already used in the ZDI work. The folding appears good, with  the unsigned $| B_{l}|$  being maximum at phase 0 and minimum at phase 0.5.

The activity indicators vary smoothly (Fig.\ref{Figure1}). $CaII~854.2$ and $H\alpha$ are synchronous, whereas $CaII~K$ deviates slightly.  Though they appear  to vary in opposition with  $| B_{l}|$, folding of the data does not fit to the photometric period, but suggests a longer period (Fig.\ref{Figure3}). We performed a period search analysis for the activity indicators $CaII~K$, $CaII~IR$ and $H\alpha$. We applied several period search techniques: Lomb-Scargle (Lomb, 1976; Scargle 1982) Lafler-Kinnman (Lafler \& Kinman 1965), Bloomfield (Bloomfield 2000), CLEANest (Foster 1995) and Deeming (Deeming 1975). Some of them gave indications of periods longer than the photometric one and mainly in the interval 10 - 15 days, but with an accuracy of 2-3 days due to the short dataset we have. 
It might be a tendency for longer periods of $CaII~IR$ and $H\alpha$, compared to $CaII~K$, but as mentioned above, we are unable to determine precise values by a period search analysis due to the short dataset. Further study on this topic with a longer dataset is required.

Taking into account possible longer rotational periods of the activity indicators and comparing them with the B$_{l}$ behaviour regarding the surface rotational period (the photometric one), there might be a correlation in their behaviour, and also a good coincidence with the ZDI map (Fig.\ref{Figure2}). The indicators have a stronger intensity at phases around $0.4 - 0.5$ when we observe areas with positive and negative polarities and the resultant B$_{l}$ is close to 0. At the opposite, they have a weaker intensity near phase 0 when the negative polarities are dominant in the resultant longitudinal magnetic field.

\section{Evolutionary considerations and dynamo operation in V390 Aur}


\subsection{Determination of stellar parameters, lithium abundance, and $v \sin i$}

\begin{figure}
   \centering
\includegraphics[width=5.2cm, angle=270, clip=true]{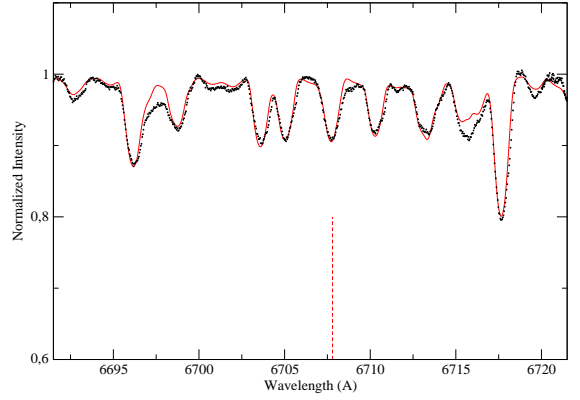}
      \caption{Fit (solid line) with the model (see text) of the spectral lines in the $Li$ region (671 nm). The dashed vertical line indicates the position of the $LiI$ 670.8 nm.}
         \label{Figure6}
   \end{figure}

With the aim of improving the precision on lithium abundance ($N_{\rm Li}=\log[n(Li)/n(H)]+12)$ and stellar parameters ($T_{\rm eff}$, $\log g$ and $[Fe/H]$), we 
performed a spectral synthesis analysis on the  Stokes I spectra of V390 Aur. We used MARCS models of atmosphere (Gustafsson et al. 2008)
and the TurboSpectrum code (Alvarez \& Plez, 1998) in order to produce high resolution synthetic spectra
of the lithium line region around 671 nm (see Canto-Martins et al. 2006 and 2011 for the complete
method and for atomic and molecular line lists references).
Best fit is displayed in Figure~\ref{Figure6}. It is obtained for $T_{\rm eff} = 4970\,K\pm50\,K$,
$\log g = 3.0\pm0.2$\,dex, and  $[Fe/H] = -0.05\pm0.05$\,dex, with a microturbulence velocity
of 1.5 \,${\rm km.s}^{-1}$\, and taking into account (with a radial tangent profile) a macroturbulence
velocity of 8 \,${\rm km.s}^{-1}$\ . Line broadening is then correctly reproduced with a rotation profile
and $v \sin i = 25\pm1\,{\rm km.s}^{-1}$\ . We checked that this solution also easily and
correctly fits other spectral regions throughout the NARVAL wavelength range.
Our  $T_{\rm eff}$ value is in perfect agreement  with that derived by Gondoin (2003) and Bell \& Gustafsson (1989) (4970$\pm$200\,K), and is very similar to the value of 5000\,K obtained by Fekel \& Balachandran (1993). 
We determined a lithium abundance $N_{\rm Li}= 1.6\pm0.15$\,dex.  This value is in very good agreement with the previous determination of $1.5\pm0.2$ made by Fekel \& Balachandran (1993), and slightly lower than the value of 1.8 obtained by Brown et al. (1989).

\subsection{Evolutionary status}

   \begin{figure}
   \centering
\includegraphics[width=8cm, angle=0]{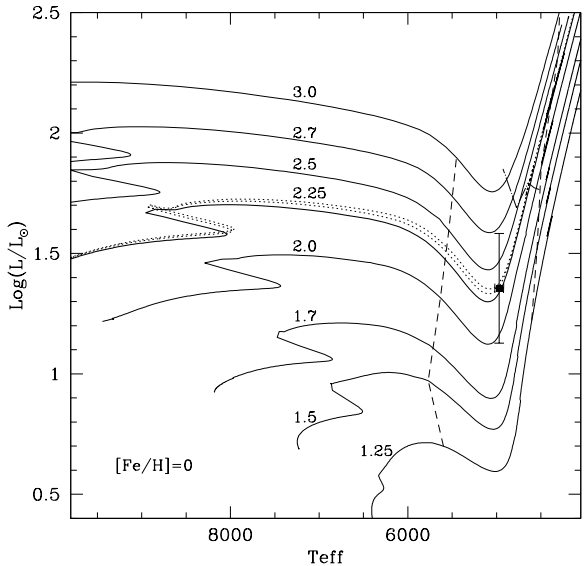}
      \caption{Position of V390 Aur in the H-R diagram. The adopted effective temperature is determined from our analysis of the Stokes I spectra of V390 Aur. The luminosity is estimated using the parallax from the New Reduction Hipparcos catalogue by van Leeuwen (2007; see text for more details). Standard evolutionary tracks (solid lines) computed for $[Fe/H]=0$ are shown up to the RGB tip for different labelled  stellar masses. For the  $2.25\,M_{\sun}$ case, the predictions for models with rotation-induced mixing treated as in Charbonnel \& Lagarde (2010) are also shown for two initial rotation velocities of 50 and 110$\mathrm{\,km.s^{-1}}$ (dotted lines). The dashed lines delimit the first dredge-up, the cooler line marking the deepest penetration of the convective envelope. The dashed-dotted line indicates the beginning of central He-burning for the mass range covered by the tracks.}
         \label{Figure5}
   \end{figure}

The position of V390 Aur in the HR diagram is shown in Fig.~\ref{Figure5}. We adopt the effective temperature of $4970\pm50\,K$ that we derived from the Stokes I spectra. 
The luminosity is obtained using the parallax of the star given in the New Reduction Hipparcos catalogue by van Leeuwen (2007), the V magnitude from the 1997 Hipparcos catalogue, and the bolometric correction following Buzzoni et al. (2010) prescription (i.e., $BC=-0.253$); the error bar reflects only the uncertainty of the parallax. The adopted parameters lead to a stellar radius of $6.4\,R_{\sun}$. 

Figure~\ref{Figure5} also presents evolutionary tracks corresponding to up-to-date stellar models of various initial masses computed with the code STAREVOL\footnote{Most of the models shown in Fig.~\ref{Figure5} are from Charbonnel \& Lagarde (2010) and Lagarde et al., in prep. The $2.25\,M_{\sun}$ models were computed especially for the present study with the same assumptions and input physics.} at solar metallicity, which is very close from the  $[Fe/H]=-0.05\pm0.05$\,dex derived from the Stokes I spectra and only slightly higher than that derived by  Fekel \& Balachandran (1993; $[Fe/H]=-0.3$).
We plot standard (non-rotating) tracks, except in the $2.25\,M_{\sun}$ case for which we also show predictions for rotating models with initial rotation velocities on the zero age main sequence of 50 and $110\mathrm{\,km.s^{-1}}$ (the higher the rotation velocity, the brighter the track). Rotating models are computed using Zahn's (1992) and Maeder \& Zahn (1998) formalism where the transport of angular momentum and chemicals is dominated by the Eddington-Sweet meridional circulation and shear instabilities (see Charbonnel \& Lagarde 2010 for details on the model assumptions and input microphysics). 
For clarity the tracks are drawn up to the RGB tip only, but the dashed-dotted line shows the location of the clump (i.e., the beginning of central He-burning) for the various masses considered.
Additionally the dashed lines indicate the beginning and the end of the first dredge-up (warmer and cooler lines respectively). 

Based on these up-to-date models and taking into account the observational uncertainties, we determine a mass of $2.25\pm0.25\,M_{\sun}$ for V390 Aur. Note that at the location of V390 Aur the effect of rotation on the mass determination is negligible compared to the uncertainties on the parallax and on the effective temperature.
The star appears to start the ascent of the Red Giant Branch (RGB) and is not bright enough to lie in the clump.
This agrees with previous studies by Gondoin (2003) and Konstantinova-Antova et al. (2009) which identified V390 Aur as a star of about $1.8-2\,M_{\sun}$ located at the base of the RGB\footnote{Charbonnel \& Balachandran (2000) determined a mass of $1.1\pm0.5\,M_{\sun}$ for V390 Aur, due to an error in the assumed Teff for this star (see their Table1). }

\subsection{Lithium abundance and rotation}

At the base of the RGB, the star experiences the so-called first dredge-up (see the dashed lines in Fig.\ref{Figure5}). Although its convective envelope has not yet reached its maximum extent in depth, it already encompasses $\sim 0.83\,M_{\sun}$ and $\sim 0.54\,R_{*}$ 
at the location in the HRD of V390 Aur. 
Starting from the lithium interstellar medium abundance of 3.3, the corresponding standard $Li$ dilution is found to lead to a theoretical surface abundance $N(Li)$ equal to 1.68 (where $N(Li)=\log[n(Li)/n(H)]+12)$ at this evolutionary point. This prediction is in rather good agreement with our observational value of 1.6$\pm$0.15. 
We note however that in this domain in stellar mass and effective temperature many subgiant and giant stars exhibit $Li$ abundances significantly lower than predicted by the standard dilution models (e.g. do Nascimento et al. 2000; de Laverny et al. 2003 ; Canto-Martins et al. 2011). This is well accounted for by additional $Li$ depletion due to rotational mixing on the main sequence (see e.g. Charbonnel \& Talon 1999, Palacios et al. 2003, Pasquini et al. 2004, Charbonnel \& Lagarde 2010, and references therein). 
For initial rotation velocities of 50 and $110\mathrm{\,km.s^{-1}}$
our $2.25\,M_{\sun}$ rotating models predict respectively  $N(Li)$ of 1.46 and 0.87 at the effective temperature of V390 Aur. The 
$Li$ content of V390 Aur can thus be explained by assuming that the star arrived on the main sequence with a modest rotation velocity of the order of $50\mathrm{\,km.s^{-1}}$.

Another possibility is that the hydrodynamical mechanisms induced by stellar rotation (i.e., meridional circulation and shear turbulence) were hampered in modifying the content of this fragile element during the evolution of the star. 
Indeed large internal angular momentum gradients, which are built up in the case of pure hydrodynamical models such as those presented in Fig.\ref{Figure5}, favour the transport of chemicals inside the star, and thus the depletion of fragile lithium as requested by $Li$ data in most field and cluster counterparts of V390 Aur (see e.g. Talon \& Charbonnel 2010 for a more detailed discussion and references).
So if an additional chief mechanism for the internal redistribution of angular momentum was able  to maintain solid body rotation, or at least to reduce strongly differential rotation in the radiative stellar interior,  much less $Li$ depletion is expected than in the pure hydrodynamical rotating case, even if the star was rotating fast at its arrival on the main sequence. 
In the mass domain of V390 Aur where the transport of angular momentum by internal gravity waves should play no role during the main sequence (Talon \& Charbonnel 2008), efficient redistribution of angular momentum, which determines the extent and magnitude of rotation-induced mixing and the resulting $Li$ depletion, could occur by hydromagnetic means related either to a fossil magnetic field or to a dynamo driven by shear in the radiative layers of the star (see e.g., Charbonneau \& MacGregor 1993, Barnes et al. 1999, Spruit 2002, Eggenberger et al. 2005, Duez et al. 2010). 
Rotating models, including the combined effects of magnetic fields, meridional circulation and shear turbulence, will be computed in the near future to quantitatively test this hypothesis in the case of stars at the evolution stage of V390 Aur. 



Additionally, let us recall that V390 Aur has a rather high rotation velocity for its evolutionary stage ($35\mathrm{\,km.s^{-1}}$ for $vsini=29\mathrm{\,km.s^{-1}}$ and $i=56\degr$), as already noted by Fekel \& Balachandran (1993). This is rather peculiar since  most (i.e., more than 95$\%$) of the stars of the same spectral class have already been spun down to a rotational velocity of $\sim 5\mathrm{\,km.s^{-1}}$ or less (Gray 1982; De Medeiros et al. 1996; De Medeiros \& Mayor 1999). This general behaviour can be explained simply by changes in the moment of inertia and in the stellar radius when the stars evolve to the RGB, and the transition to a slow rotation occurs at an effective temperature at the order of 5500K. Taking into account this secular effect and assuming rigid-body rotation and conservation of angular momentum for our $2.25\,M_{\sun}$ model, one can obtain a maximum rotation velocity of only $\sim15\mathrm{\,km.s^{-1}}$ (for a rotation velocity of $180\mathrm{\,km.s^{-1}}$ on the zero age main sequence) at the location of V390 Aur in the HRD.
 
Although fast rotation is observed only in few percents of red giant stars, it appears to be a common feature of a group of chromospherically active single giants (Fekel \& Balachandran, 1993; Konstantinova-Antova et al. 2009). For the moment, possible explanations for their faster rotation could be angular momentum dredge-up from the faster rotating interior during the first dredge-up phase (Simon \& Drake 1989), or spin-up from planet engulfment that could also increase the lithium abundance (e.g. Siess \& Livio 1999; Livio \& Soker 2002; Massarotti et al. 2008; Carlberg et al. 2009). 
The first option requires rotation gradient in the stellar interior with the near core region spinning much faster than the more external layers. According to the theoretical works by Palacios et al. (2006) and Brun \& Palacios (2009) rotational gradients, both radial and longitudinal ones, should appear in the convective envelope and the radiative zone for stars with mass up to $2-2.5\,M_{\sun}$ during the first dredge-up phase and the beginning ascend of the RGB. Our observations support their predictions for gradients in the convective envelope. In addition, a gradient between the envelope and (presumable) fast rotating radiative interior could dredge--up angular momentum and speed-up the upper convective layers and the surface (as we observe in V390 Aur and other fast rotating single giants). 

However, one should be cautious with all these interpretations, since the knowledge on the various transport mechanisms during the RGB phase is not complete yet. It is rather possible, not only meridional circulation and shear-induced turbulence to play role for transport of angular momentum and chemicals, but also other factors, like magnetic fields etc. Further work in this direction is required. 

On the other hand, recent discoveries of Jupiter-mass planets orbiting their host star at inner-solar system distances could also be considered as a possible explanation of the $Li$ content and fast rotation in V390 Aur. Applying the formula by Massarotti (2008) with the parameters obtained by our model and assuming a circular orbit with a semi-major axis equal to the radius of the star (hypothesis for the just engulfed planet, that could explain not only the fast rotation, but also the enchanced $Li$), we estimate that the star has to engulf a brown dwarf of about $20M_{Jup}$ to explain its $V_{eq}$ of $34\mathrm{\,km.s^{-1}}$. In the same time, however, Alibert et al. (2011) point out the observed lack of short period planets orbiting $2\,M_{\sun}$ stars.

As to the forcing of stellar rotation by a close companion, let us recall that V390 Aur was found to be a wide binary system and synchronization cannot play a role in its fast rotation (Konstantinova-Antova et al. 2008).

\subsection{Dynamo}
On the basis of the theoretical convective turnover time and assuming that the photometric period (9.825 d) is the rotational one, we can estimate the value of the Rossby number for V390 Aur. If the value of 126 days (for a layer that corresponds to $1/2H_{p}$ above the base of the convective envelope) is considered, which is obtained from our standard $2.25\,M_{\sun}$ model (computed with a mixing length parameter of 1.6) at the location of V390 Aur in the HR diagram, then the Rossby number is 0.08. In the case of convective turnover time of 52 days, i.e. the value determined for the middle (in radius) of the convective envelope, the Rossby number is 0.19. In both cases, these values are indicative of a very efficient $\alpha$--$\omega$ dynamo.

\section{Discussion}

The present study revealed a complex structure of the magnetic field at the surface of V390 Aur. In addition to the magnetic polar spot of positive polarity, groups of close situated magnetic spots of negative polarity are detected at mid-latitudes. Such polar spots are predicted for fast rotators by Sch\"ussler (1996), and observed by Petit et al. (2004). Temperature polar spots are reported for giants in RS CVn systems and for FK Com--type stars by Strassmeier (1997), Korhonen et al. (2002) and Strassmeier (2002). 

This surface structure with close magnetic spots could explain the flare activity properties of the star reported in optical and X--ray (Konstantinova-Antova et al. 2000; Gondoin 2003; Konstantinova-Antova et al. 2005), like the groups of optical flares and the "continuous flare activity" in the corona. 

The observed azimuthal map belt at high-latitude could be interpreted as a toroidal component near the surface (like in the FK Com--type star HD 199178 and in the RS CVn star HR 1099, Petit et al. 2004a and b) and is indicative of the site of the dynamo operation. In this star, it is possible for the dynamo to operate not close to the base of the convective envelope, but in a whole region within it. The other possibility is the toroidal magnetic field to be formed near the base of the convection zone and to rise up as one whole. This possibility is less likely, because of the vigorous convection during the first dredge-up phase. The first assumption, however, presumes the existence of a gradient of rotation in the convective envelope closer to the surface, not only at its base. The fast surface rotation of V390 Aur and the significant differential rotation $\Delta\Omega=0.048\mathrm{\,rad.d^{-1}}$ are in support of such a rotational gradient.

The comparison to solar-type dwarfs revealed that Sun-like stars seem to display a higher fraction of toroidal field for a rotation period similar to that of V390 Aur (Petit et al. 2008). The mostly poloidal field geometry is more reminiscent of M dwarfs with deep convective envelopes (Morin et al. 2008, 2010). But, one must be cautious about this question, because the fraction of toroidal field can vary significantly during stellar cycles (Petit et al. 2009).

\section{Conclusions}

\begin{enumerate}
  \item The complex structure of the surface magnetic field in the ZDI map and the analysis of the poloidal and toroidal field components suggest a dynamo operation in V390 Aur. It is rather different than the ZDI map of another late G giant, EK Eri, a descendant of an Ap star, where a strong dominating dipole structure is observed (Auriere et al. 2011). In addition, it is rather likely for the dynamo to be of $\alpha$--$\omega$ type, taking into account the fast rotation and the evolutionary stage of V390 Aur with a well-developed convective envelope. The calculated Rossby number obtained from tailor-made stellar evolution models is also in support of an efficient $\alpha$--$\omega$ dynamo. A significant toroidal component is developed which results in low-latitude spots and an azimuthal field belt at the surface. Such a belt is already observed in FK Com and RS CVn - type stars. A possible explanation is that there is a gradient of the rotation in the extended convective envelope which exists during the first dredge-up phase and it is possible for the dynamo to operate within it, not only at the base of this envelope, as it is the case of a classical solar-type dynamo, but also closer to the surface. 
   \item The structure of the surface magnetic field with close situated active areas could also explain the properties of the flare activity in V390 Aur, observed in the past: groups of optical flares and continuous flaring in the corona.
    \item The activity indicators behavior points to a possible decrease of the rotation rate in the chromosphere compared to the rate in the photosphere. Further study is necessary to determine the exact rotation periods for the activity indicators in the chromosphere.
 
    In the future, it would be interesting to study more single giants - their surface magnetic field configurations and the dynamo, and to get knowledge of the structure of their atmospheres, using ZDI. Recently, we started a ZDI programme of eight giants with rotational periods from 5 days to 600 days, including stars from the Hertzsprung gap to the clump. In the future, it would be also possible for ZDI to be applied to AGB stars. Even if for some of these giants small $vsini$ do not allow good resolution mapping, such mapping makes it possible to determine the surface magnetic field strength, to disentangle poloidal and toroidal components and finally to make diagrams similar to those obtained for M dwarves and solar twins (Morin et al. 2010). 
\end{enumerate}

\begin{acknowledgements}
We thank the TBL team for providing service mode observing with NARVAL. R.K.- A. is thankful to Dr. Sara Palmerini for a useful discussion. The observations were funded under an OPTICON grant in 2008. R.K.-A. is thankful for the possibility to work for six months in 2010 as a visiting researcher in LATT, Tarbes under Bulgarian NSF grant DSAB 02/3/2010. R.K.- A., S.Ts. and R.B. acknowledge partial financial support under NSF contract DO 02-85. R.K.-A. and R.B. also acknowledges support under the RILA/EGIDE exchange program (contract RILA 05/10). C.C. aknowledges financial support from the Swiss National Science Foundation (FNS) and the French Programme National de Physique Stellaire (PNPS) of CNRS/INSU.
\end{acknowledgements}

\end{document}